\documentclass[prl, twocolumn,preprintnumbers,amsmath,amssymbl]{revtex4-1}

\usepackage{graphicx}
\usepackage{dcolumn}
\usepackage{bm}
\usepackage{subfigure}

\begin{document}\
\author{S.~P.~D.~Mangles$^1$}
\author{G.~Genoud$^2$}
\author{M.~S.~Bloom$^1$}
\author{M.~Burza$^2$}
\author{Z.~Najmudin$^1$}
\author{A.~Persson$^2$}
\author{K.~Svensson$^2$}
\author{A.~G.~R.~Thomas$^3$}
\author{C.-G.~Wahlstr\"om$^2$}

\title{The self-injection threshold in self-guided laser wakefield accelerators}
\affiliation{$^1$The Blackett Laboratory, Imperial College London, SW7 2AZ, UK}
\affiliation{$^2$Department of Physics, Lund University, P.O. Box 118, S-22100 Lund, Sweden}
\affiliation{$^3$Center for Ultrafast Optical Science, University of Michigan Ann Arbor, Michigan 48109 USA}

\begin{abstract}
A laser pulse traveling through a plasma can excite large amplitude plasma waves that can be used to accelerate relativistic electron beams in a very short distance---a technique called laser wakefield acceleration.  
Many wakefield acceleration experiments rely on the process of wavebreaking, or self-injection, to inject electrons into the wave, while other injection techniques rely on operation without self-injection.
We present an experimental study into the parameters, including the pulse energy, focal spot quality and pulse power, that determine whether or not a wakefield accelerator will self-inject.
By taking into account the processes of self-focusing and pulse compression we are able to extend a previously  described theoretical model, where the minimum bubble size $k_pr_b$ required for trapping is not constant but varies slowly with density and find excellent agreement with this model.  
\end{abstract}

\maketitle

Laser wakefield acceleration, where an intense laser pulse drives a plasma wave with a relativistic phase velocity, is a promising technique for  the development of compact, or ``table-top'', particle accelerators and radiation sources.  
Plasma waves driven in moderate density plasmas can support electric fields over a thousand times stronger than those in conventional accelerators.
Laser driven plasma waves have demonstrated electron acceleration to $\simeq 1$ GeV in distances $\simeq$~1~cm \cite{LeemansWP_NaturePhysics_2006, KneipS_PRL_2009, Clayton_PRL_2010}. 
These compact particle accelerators have significant potential as bright x-ray sources \cite{RousseA_PRL_2004, KneipS_NaturePhysics_2010,  FuchsM_NaturePhysics_2009} offering peak brightness comparable  to 3rd generation synchrotron sources in x-ray flashes on the order of just 10 fs.  

At the heart of the laser wakefield acceleration concept is the fact that electron plasma waves with relativistic phase velocities are driven to very large amplitudes, where they become highly non-linear.  
If the plasma wave is driven beyond a threshold amplitude, the wave breaks.  
When the wave is driven far beyond the wavebreaking threshold, the wave structure is destroyed and large amounts of charge can be accelerated to high energy but with a broad energy spread \cite{ModenaA_Nature_1995}.  
With appropriately shaped laser pulses this normally catastrophic process of wavebreaking can be tamed to produce high quality beams of electrons.
This is because close to the wavebreaking threshold the nature of wavebreaking changes -- some electrons from the background plasma can become trapped in the wave without destroying the wave structure, a process called self-injection.

The highly non-linear broken wave regime \cite{PukhovA_APB_2002} is used in many experiments  to produce quasi-monoenergetic electron beams \cite{ManglesSPD_Nature_2004, GeddesCGR_Nature_2004, FaureJ_Nature_2004}.   
In such experiments a threshold plasma density is commonly observed, below which no electron beams are produced.  
Due to the inverse scaling of the electron beam energy with plasma density, the highest energy beams achievable with a given laser system are achieved just above the threshold, and it is well known that many of the beam parameters including the spectrum and stability are also optimised just above the threshold density \cite{MalkaV_PoP_2005, ManglesSPD_PoP_2007}.
It is also well known that to achieve self-injection at lower densities higher power lasers are required - although the exact scaling of the threshold with laser power is not well known.
A number of techniques to improve the electron beam parameters including stability and total charge, have recently been demonstrated by using alternative injection schemes \cite{FaureJ_Nature_2006, McGuffeyC_PRL_2010, PakA_PRL_2010, GeddesCGR_PRL_2008, Schmid_PRSTAB_2010}. 
Crucially these schemes all rely on operating the LWFA below the self-injection threshold.
A number of recent purely theoretical papers have addressed the dynamics of wavebreaking or self-injection \cite{KalmykovS_PRL_2009, KostyukovI_PRL_2009, ThomasAGR_PoP_2010, Yi_SA_PPCF_2011}.
Clearly a good understanding of the self-injection threshold is important for the development of laser wakefield accelerators.
We report here on a series of experiments which identify the key laser and plasma parameters needed to predict the density threshold and we develop a model capable of predicting the self-injection threshold density for a given set of experimental parameters.

In LWFA experiments the laser pulse self-focuses due to the transverse non-linear  refractive index gradient of the plasma  \cite{ThomasAGR_PRL_2007, RalphJE_PRL_2010} and the spot size decreases towards a matched spot size.  
This matched spot size occurs when the ponderomotive force of the laser balances the space charge force of the plasma bubble formed.
In situations where there is no loss of energy during self-focusing, nor any change in the pulse duration,the final matched spot size, and hence the final intensity is simply a function of $\alpha P/P_c$.
$P$ is the laser power; $\alpha$ is the fraction of laser energy within the full width at half maximum intensity of the focal spot---important because energy in the wings of the spot are not self-focused by the plasma wave and so do not contribute; 
$P_c$ is the laser power where  relativistic self-focussing dominates over diffraction, $P_c= (8\pi\epsilon_0m_e^2c^5/e^2)(n_c/n_e) \simeq 17n_c/n_e$~GW (where $n_e$ is the background plasma electron density and $n_c$ is the critical density for propagation of the laser in the plasma).
We might therefore expect that the self-injection threshold would occur at a fixed value of $\alpha P/P_c$ \cite{FroulaDH_PRL_2009}.  
However it is also known that the longitudinal non-linear refractive index gradient also has a significant effect on the pulse properties \cite{FaureJ_PRL_2005, SchreiberJ_PRL_2010} and we expect this to have an affect on the self-injection threshold.

The experiment was carried out using the multi-TW laser at the Lund Laser Centre.  
The laser delivered pulse energies of up to 0.7 J in pulses as short as 40~fs, corresponding to a peak power of  18 TW.  
An f/9 off-axis parabolic mirror was used to  focus the pulse. 
A deformable mirror was used to optimise the focal spot, producing a spot size of $16 \pm 1~\mu$m \textsc{fwhm}.
For a gaussian focal spot the theoretical maximum fraction of energy within the \textsc{fwhm} is $\alpha = 1/2$, the best focus that we obtained had $\alpha = 0.48$.
The focal plane was positioned onto the front edge of a supersonic helium gas jet with an approximately flat-top profile of length $1.8 \pm 0.1$~mm.   

To investigate the self-injection threshold we studied the effect of the plasma density, $n_e$, the total laser energy, $E$, the focal spot quality, $\alpha$, and the pulse duration, $\tau$, on the amount of charge in the electron beam.
We chose to use the total charge in the electron beam as the diagnostic of self-injection as it provides a clear unambiguous signal of an electron beam.

The charge was measured using an electron beam profile monitor, consisting of  a lanex  screen placed on the back surface of a wedge (which was used to collect the transmitted laser light).  
The wedge was 1~cm thick and made of glass and therefore prevented electrons below approximately 4 MeV reaching the lanex.  
The lanex screen was imaged onto a 12 bit \textsc{ccd} camera.
To reduce the amount of background light from the interaction, a narrow band interference filter matched to the peak emission of the lanex screen was placed in front of the camera. In addition the camera was triggered several microseconds after the interaction but within the lifetime of the lanex fluorescence. 
The lanex screen was calibrated using the absolute efficiency data, absolute response of the \textsc{ccd} camera and the details of the imaging system \cite{Glinec_RSI_2006}
A beam profile monitor was used in preference to an electron spectrometer due to the fact that it has a higher sensitivity (i.e. the signal produced by a low charge beam dispersed inside a spectrometer will drop below the background level, whereas the same low charge beam will produce a bright image on the profile monitor). 
Also close to the threshold we do not expect the electrons to have particularly high energy (i.e. injection could be occurring but the electron beam energy could be outside the range of the electron spectrometer). .

The gas jet could produce electron densities up to $ n_e =  5 \times 10^{19}$~cm$^{-3}$.
The laser pulse energy was varied by altering the energy pumping the final laser amplifier. 
We used the deformable mirror to reduce $\alpha$ by adding  varying amounts of spherical aberration. 
Spherical aberration has the effect of decreasing $\alpha$ without introducing asymmetry to the focal spot and without significantly affecting its size.
Degrading the focal spot symmetrically was desirable as asymmetric pulses can drive asymmetric wakes which can have a strong effect on the dynamics of self-injection \cite{ManglesSPD_APL_2009}. 
 The pulse duration was altered by changing the separation of the gratings in the compressor.  
Changing the grating separation  introduced both a chirp to the pulse spectrum and a skew to the pulse envelope.
To take this into account we investigated both positive and negative chirps. 

\begin{figure}
\begin{center}
\includegraphics[width=8.5cm]{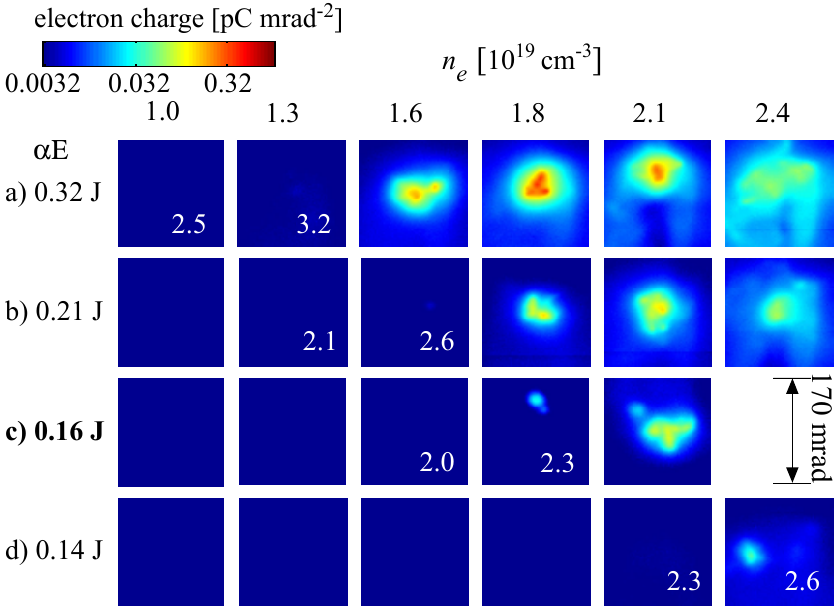}
\caption{(colour online) Electron beam profiles for various plasma densities for different values of the amount of laser energy within the \textsc{fwhm} of the focal spot.  a), b) and d) kept the total laser energy constant but varied $\alpha$ whereas c) reduced the laser energy. 
Each panel is an average of 5 shots and is displayed on a logarithmic colourscale}
\label{alphaE}
\end{center}
\end{figure}

Figure \ref{alphaE} shows the effect of varying the laser pulse energy within the focal spot on the self-injection threshold.  
Keeping the total laser energy constant and degrading the focal spot (i.e. lowering $\alpha$) moves the threshold to higher plasma densities.  
We also observe an increase in the threshold density when we keep  $\alpha$ constant and reduce the laser pulse energy.
In fact we find that the two effects are equivalent, i.e. that the threshold shifts according to the product $\alpha E$.  
This demonstrates that it is only the energy within the \textsc{fwhm} of the focal spot that contributes to driving the plasma wave.  
This emphasises the importance of laser focal spot quality in LWFA experiments \cite{IbbotsonTPA_PRSTAB_2010}, which are often performed with $\alpha \approx 0.3$ \cite{KneipS_PRL_2009, FroulaDH_PRL_2009}.  
Improving the focal spot could therefore result in a significant increase in the electron beam energy achievable from a given laser system.

The observed variation of the threshold  with $\alpha E$ is as expected for one based on $\alpha P/P_c$ but this can only be confirmed by the behaviour of the threshold when we vary the laser pulse duration, keeping $\alpha E$ constant.
When we do this we see markedly different behaviour.

\begin{figure}
\begin{center}
\includegraphics[width=8.5cm]{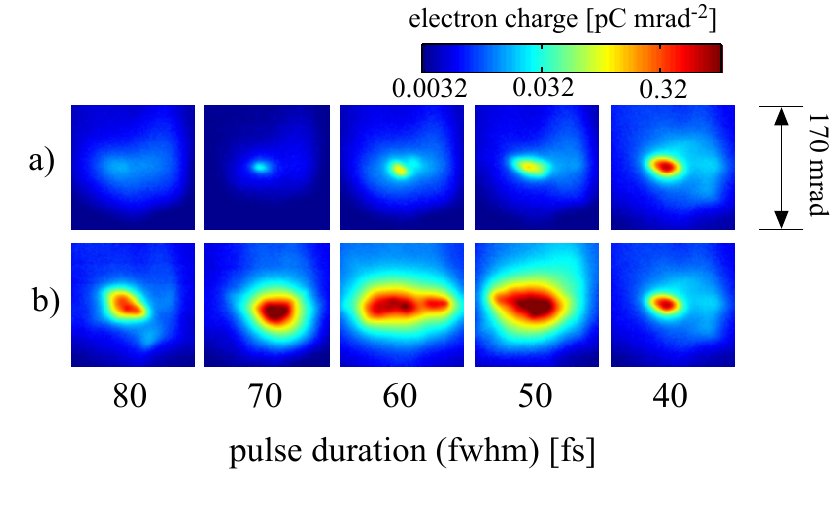}
\caption{Electron beam profiles for various pulse durations at fixed $\alpha E$ and at a plasma density just above the threshold density for injection for 40 fs pulses.  
The pulse duration was varied by changing the compressor grating separation which introduces a chirp to the pulse a) negative chirp  b) positive chirp }
\label{duration_scan}
\end{center}
\end{figure}

We kept the plasma density constant, at a value just above the threshold density for the optimally compressed pulse. 
At this density ($n_e = 1.6 \times 10^{19}$~cm$^{-3}$), with full laser energy ($\alpha E = 0.32$~J) and the fully compressed pulse ($\tau  = 42$~fs) we observed a bright electron beam.
When we reduced either the plasma density or the pulse energy by a small factor (20 - 25\%) this beam disappeared, i.e. we dropped below the threshold. 
Even after increasing the pulse duration by a factor of two electrons are clearly still injected, as shown in fig \ref{duration_scan}.  
This is true regardless of the chirp of the laser pulse, however we do see an enhancement of the total charge using positively chirped (red at the front) pulses as reported previously  \cite{LeemansW_PRL_2002}.
These pulses have a fast rising edge indicating that the precise shape of the pulse may play a role in the total charge injected. 
The direction of chirp of the pulse may also affect the rate at which pulse compression occurs \cite{Dodd_POP_2001}.
For both directions of chirp the fact that the threshold behaviour is so significantly different to that observed when varying $\alpha E$  suggests  that pulse compression is indeed playing an important role in determining whether or not the accelerator reaches wavebreaking.

\begin{figure}
\begin{center}
\includegraphics[width=8.5cm]{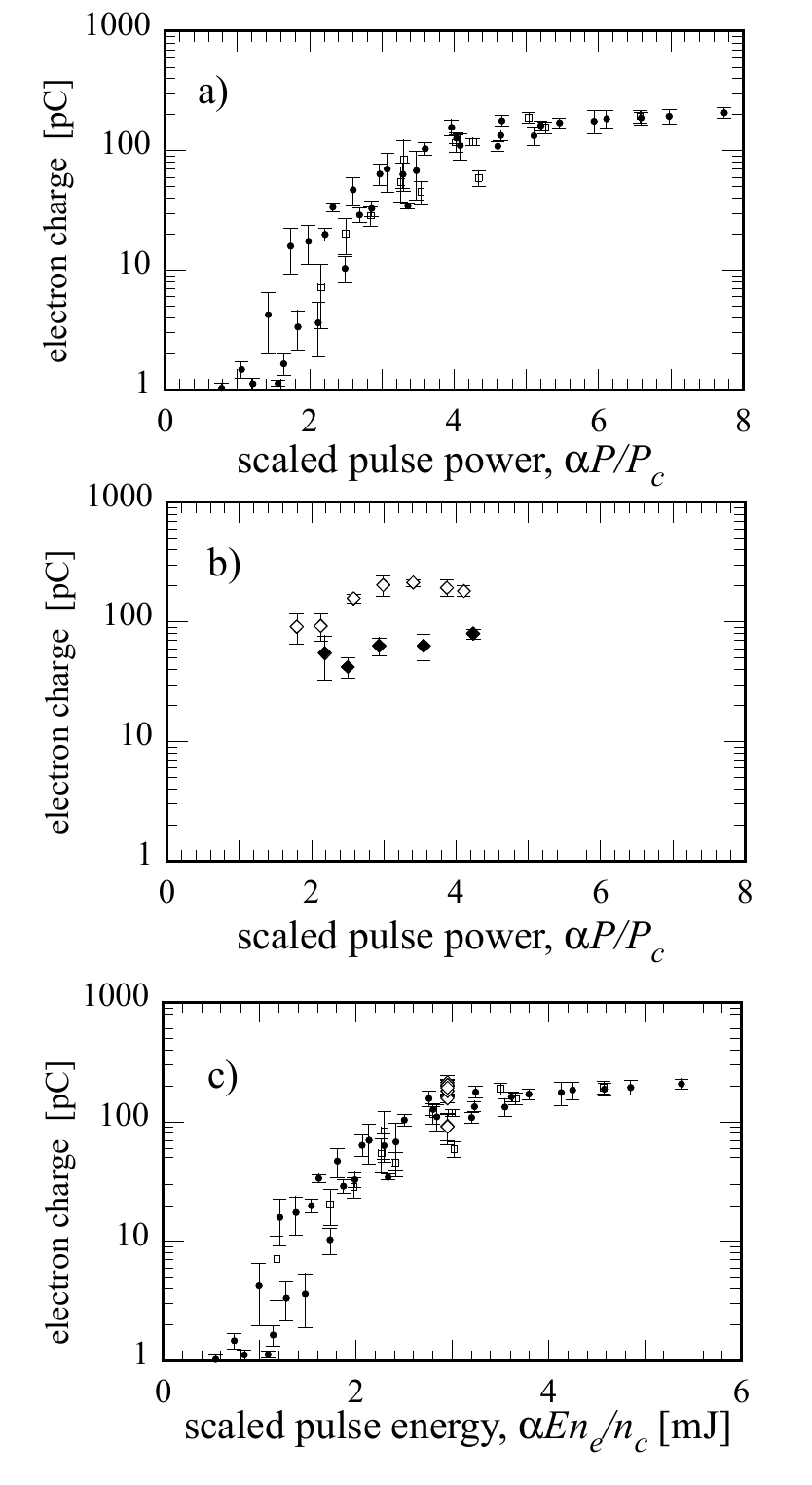}
\caption{a) electron charge  ($>$~4~MeV) versus $\alpha P/P_c$ keeping the pulse duration constant but varying focal spot quality and plasma density (closed circles) or total pulse energy and plasma density (open squares) but keeping pulse duration constant.  
b) electron charge versus $\alpha P/P_c$ varying pulse duration while keeping plasma density and energy in focal spot constant.
c) data from (a) and (b) plotted versus  $\alpha En_e/n_c$ 
Each data point is an average of five shots and the error bars represent one standard deviation.}
\label{adjusted_energy5}
\end{center}
\end{figure}

In figure \ref{adjusted_energy5} we plot the total charge observed on the profile monitor screen for the various data sets.  
Figure \ref{adjusted_energy5}a) shows the total charge, plotted against the pulse power normalised to the critical power for self-focusing, for the data sets where we varied the plasma density and the energy within the focal spot (either by varying the spot quality $\alpha$ or total pulse energy $E$). 
The charge rises rapidly with increasing $\alpha P/P_c$ until eventually reaching a plateau at around $\alpha P/P_c \approx 4$.  
There is an increase in the total charge of  a factor of ten between $\alpha P/P_c = 2$ and $\alpha P/P_c = 4$ for both sets of data.  
The fact that both datasets lie on the same curve confirms the fact that it is the energy within the focal spot which determines the wakefield behaviour. 
This supports the hypothesis that energy in the wings of the focal spot is not coupled into the accelerator: energy in the wings of the spot is effectively wasted.

Figure  \ref{adjusted_energy5}b) shows the charge plotted against $\alpha P/P_c$ for a data set where we kept the plasma density and  $\alpha E$ constant but varied the pulse duration (by introducing either positive or negative chirp).
The markedly different behaviour is once again apparent: rather than the rapid increase of charge between  $\alpha P/P_c = 2$ and $\alpha P/P_c = 4$ the charge is approximately constant for each data set.

Figure  \ref{adjusted_energy5}c) plots all of the data sets (varying $\alpha$, $E$ and $\tau$)  against a scaled pulse energy  $\alpha En_e/n_c$  rather than the scaled pulse power.  
The fact that the pulse duration dataset now fits  closely with the $\alpha E$ datasets confirms that pulse compression is playing an important role in determining whether or not the wakefield accelerator reaches self-injection.

A recent paper that examined the trajectory of electrons inside the plasma bubble \cite{ThomasAGR_PoP_2010} predicts that self-trapping will occur when the radius of the plasma bubble ($r_b$) is larger than a certain value given by:
\begin{equation}
\label{AGRT_threshold}
k_pr_b > 2\sqrt{\ln(2\gamma_p^2)-1}
\end{equation}
Where  $\gamma_p  \approx \sqrt{n_c/(3n_e)}$~\cite{DeckerCD_PoP_1996} is the Lorentz factor associated with the phase velocity of the bubble.
When this condition is met, an electron starting at rest a distance $r_b$ from the laser axis and following an elliptical trajectory in the bubble fields (thus defining the edge of the bubble) will be accelerated by the bubble fields up to $\gamma_p m_ec^2$ by the time it reaches the back of the bubble.
A key feature of this model is that the normalised bubble size required for self-injection $k_pr_b$ is not constant with density.
As equation \ref{AGRT_threshold} depends only on the plasma density and bubble size we can determine the minimum pulse properties required to reach the threshold by noting that the radius of the bubble is related to the pulse energy and duration through \cite{LuW_PRSTAAB_2007}:
\begin{equation}
\label{kprb}
 k_p r_b  =2\sqrt 2\left(\frac{\alpha E}{\tau P_c}\right)^\frac{1}{6}
\end{equation}
Combining equations \ref{AGRT_threshold} and \ref{kprb} yields an expression for the minimum pulse energy required to reach  self-injection:
\begin{equation}
\label{energy:thresh}
\alpha E   >  \frac{\pi\epsilon_0m_e^2c^5}{e^2} \left[\ln\left({\frac{2n_c}{3n_e}}\right)-1\right]^3 \frac{n_c}{n_e}\tau(l)
\end{equation}
where $\tau(l)$ is the pulse duration after a propagation length $l$.  
A simple model for the rate of pulse compression was put forward in ref. \cite{SchreiberJ_PRL_2010} based on the fact that the front of the pulse travels at the group velocity of the laser in the plasma and the back of the pulse travels in vacuum, this produces $\tau(l) \approx \tau_0 -  (n_e l)/(2cn_c)$.
The interaction length will be limited by either the length of the plasma target or the pump depletion length $ l_{pd} \simeq c\tau_0n_c/n_e$ \cite{LuW_PRSTAAB_2007}.
For the depletion limited case equation \ref{energy:thresh} reduces to:
\begin{equation}
\label{pump_depletion_version}
\frac{\alpha P}{P_c}>  \frac{1}{16} \left[\ln\left({\frac{2n_c}{3n_e}}\right)-1\right]^3 
\end{equation}
The threshold density for self-injection for a given experiment can be calculated from   \ref{energy:thresh} and \ref{pump_depletion_version}.
This model requires knowledge of the initial pulse energy, pulse duration and the length of the plasma to predict the threshold. 
As equations \ref{energy:thresh} and \ref{pump_depletion_version} are transcendental,  the density threshold for a given laser system must be found numerically.

A previous study showed that, at low density, the threshold is approximately $\alpha P/P_c > 3$ \cite{FroulaDH_PRL_2009},  this can be rearranged into a similar form to equation \ref{energy:thresh}:
\begin{equation}
\label{P_Pc3}
\alpha E > 3 \frac{\pi\epsilon_0m_e^2c^5}{e^2} \frac{n_c}{n_e}\tau_0
\end{equation}
We can then use equation \ref{P_Pc3} to predict  the density threshold for specific experimental conditions.  
To use this model only the initial pulse power is required to calculate the threshold density.
Combining $\alpha P/P_c > 3$ and equation \ref{kprb} reveals that this threshold model  is also equivalent to stating that the minimum bubble size for self-trapping is constant with density ($k_pr_b > 3.4$) in contrast to equation \ref{AGRT_threshold}.

\begin{figure}
\begin{center}
\includegraphics[width=8.5cm]{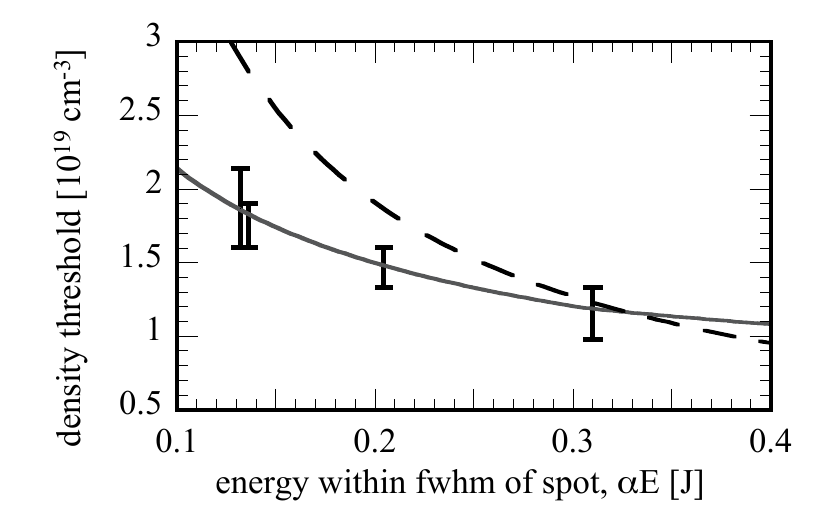}
\caption{Observed density threshold as a function of laser energy ($\alpha E$) for our experiment.   
The solid curve represents our threshold model. 
The dashed curve represents a threshold based on $\alpha P/P_c > 3$.
}
\label{figure4a}
\end{center}
\end{figure}

In figure \ref{figure4a} we plot the variation of the observed threshold density with laser energy ($\alpha E$).  
We have defined the  experimentally observed threshold density as lying in the region between the highest density where we observe no electron beam and the lowest density where we clearly observe a beam.  
We also show the theoretical threshold density based on equations \ref{energy:thresh} and \ref{pump_depletion_version}, and the predicted threshold based on equation \ref{P_Pc3}.
Its agreement with the experimental data indicates that our model accurately predicts the self-injection threshold, confirming that the threshold is reached because the laser pulse undergoes intensity amplification due to a combination of pulse compression and self-focusing.


Our measurements of the threshold density for self-injection have been made with only moderate laser pulse energies $\sim1$~J.
Many laser wakefield experiments are now being performed with pulse energies $\sim$~10~J and the validity of this model at these higher laser energies can be verified by applying it to previously published data.   
We restrict ourselves to data obtained from experiments with gas-jets as guiding structures can affect the trapping threshold by changing the way pulse evolution occurs \cite{IbbotsonTPA_PRSTAB_2010} or by introducing additional effects such as ionization injection \cite{RowlandsReesTP_PRL_2008}.
To calculate the density threshold for a particular set of experimental parameters the following information is required: the laser energy $E$, the focal spot quality $\alpha$, the initial pulse duration $\tau$ and the maximum plasma length $l$.  
Equations \ref{energy:thresh} and \ref{pump_depletion_version} or equation \ref{P_Pc3} can then be used to calculate the expected density threshold for the two models.  
Kneip \emph{et al} \cite{KneipS_PRL_2009}, using a 10 J,  45~fs, 800~nm laser pulse with  $\alpha = 0.3$, observed a threshold density of $n_e =  2 - 3 \times 10^{18}$~cm$^{-3}$ in an 8.5 mm long plasma; 
our model predicts that the threshold density for self-injection should occur at $n_e \approx 3 \times 10^{18}$~cm$^{-3}$.
Froula \emph{et al} \cite{FroulaDH_PRL_2009}, using  a 60~fs, 800~nm laser with $\alpha E \approx 6$~J observed a threshold density of  $n_e \approx 3 \times 10^{18}$~cm$^{-3}$ in an 8.0~mm plasma; our model also predicts $n_e \approx 3 \times 10^{18}$~cm$^{-3}$.  
Schmid \emph{et al} \cite{SchmidK_PRL_2009} using an 8~fs, 840~nm laser with $\alpha E \approx 15$~mJ observed electron beams at a density of $n_e \approx 2 \times 10^{19}$~cm$^{-3}$ in  a plasma 300~$\mu$m long; our model predicts a threshold of 
$n_e \approx 2.2 \times 10^{19}$~cm$^{-3}$. 
Faure \emph{et al} \cite{FaureJ_Nature_2004}, using a 33~fs, 820~mn laser, reported a dramatic decrease in the number of accelerated electrons at $n_e \approx 6 \times 10^{18}$~cm$^{-3}$ in a 3~mm gas jet with $\alpha E \approx 0.5$~J.  
Our model predicts a threshold density of  $n_e \approx 7 \times 10^{18}$~cm$^{-3}$.

\begin{figure}
\begin{center}
\includegraphics[width=8.5cm]{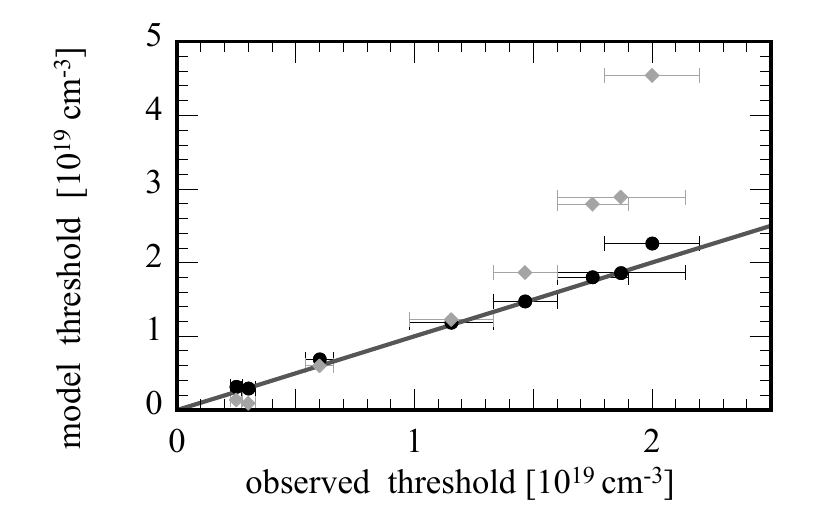}
\caption{Plot of reported density threshold, $n_t$, versus predicted density threshold, $n_{\rm model}$, for this and other published experiments \cite{KneipS_PRL_2009, FroulaDH_PRL_2009, FaureJ_Nature_2004, SchmidK_PRL_2009}.  
Circles show the predictions of our model, Diamonds show the threshold based on $\alpha P/P_c > 3$.
The line indicates $n_t = n_{\rm model}$.}
\label{figure4b}
\end{center}
\end{figure}

These additional data points, together with those from this experiment are presented in figure \ref{figure4b}.
Due to the fact that our model does not depend on a single experimental parameter we plot the experimentally observed density threshold $n_t$ for each experiment on the $x$-axis and against the calculated threshold $n_{\rm model}$  obtained using either equations  \ref{energy:thresh} and \ref{pump_depletion_version} or equation \ref{P_Pc3}.   
Figure \ref{figure4b} shows that our model is in good agreement with experiments over nearly three orders of magnitude in laser energy, whereas the threshold based on equation \ref{P_Pc3} matches the observed threshold over only a very limited range of pulse energies: it overestimates the threshold density for  low energy  laser systems and on the other hand would significantly underestimate the threshold for very high energy laser systems.

We note that simulations by Yi \emph{et al}., \cite{Yi_SA_PPCF_2011} show that, at very low density and an initial laser spot size less than the matched spot size, diffraction of the laser pulse leads to a lengthening of the bubble which plays a role in determining self-injection.  
In that work they see self-injection with a 200 J, 150 fs laser pulse at a density of $n_e = 10^{17}$~cm$^{-3}$. 
Our model predicts that the threshold would be $n_e \approx 4 \times 10^{17}$~cm$^{-3}$---actually in reasonable agreement with  \cite{Yi_SA_PPCF_2011}, however our model relies on pulse compression occurring over $\approx 10$~cm whereas Yi \emph{et al}., show that in their simulations injection occurs after just 5 mm .  
This indicates that  our model is only  valid for initial laser spot sizes \emph{greater than or equal to} the matched spot size (as is the case for  the experiments shown in figure \ref{figure4b}).

We now use our model to predict the self-injection threshold density for lasers currently under construction. 
For example, our model predicts that a 10 PW laser (300~J in 30~fs, $\lambda = 0.9~\mu$m, such as the Vulcan 10~PW laser at the Rutherford Appleton Lab, or the ELI Beamlines facility in the Czech Republic) could produce electron injection at as low as $n_e \approx 2 \times 10^{17}$~cm$^{-3}$ (assuming $\alpha = 0.5$)  in a 6~cm long plasma. 
For  a 1 PW  laser (40~J in 40 fs, $\lambda = 0.8~\mu$m, such as the Berkley Lab Laser Accelerator, BELLA) our model predicts that self-injection will occur at a density of $n_e \approx 9 \times 10^{17}$~cm$^{-3}$ in 2.4 cm.

The lower the threshold density of a wakefield accelerator, the higher  the maximum beam energy.  
However, for self-injecting accelerators there must be acceleration after injection, requiring operation at densities slightly above this threshold so that injection occurs earlier in the interaction.  

In summary, we have measured the effect of various laser parameters on the self-injection threshold in laser wakefield accelerators.  
The simple model we use relies on the fact that pulse compression and self-focusing occur and that only the energy within the \textsc{fwhm} of the focal spot  contributes towards driving the plasma wave. 
We find that in cases where the interaction is limited by pump depletion, the threshold can be expressed as a ratio of $P/P_c$,  but this ratio is not the same for all laser systems:  for higher power lasers the threshold occurs at a higher value of $P/P_c$ than for lower power lasers. 
When the plasma length is shorter than the pump depletion length we find that the length of the plasma is an important parameter in determining the injection threshold.

This work was supported by the Royal Society, EPSRC (Grant number EP/I014462/1);  the Swedish Research Council (including the Linn\'e grant to the LLC); the Knut and Alice Wallenberg Foundation; the EU Access to Research Infrastructures activity, FP7 grant agreement No 228334: Laserlab Europe, the Lund University X-ray Centre (LUXC); and the Marie Curie Early Stage Training Site MAXLAS (Contract No. MEST-CT-2005-020356).

\end{document}